\documentclass[11pt]{article} 

\makeatletter\renewcommand{\section}{\@startsection
{section}{1}{\z@}{-3.5ex plus -1ex minus
    -.2ex}{2.3ex plus .2ex}{\bf }}

\makeatletter\renewcommand{\subsection}{\@startsection{subsection}{2}{\z@}{-3.25ex
plus -1ex minus
   -.2ex}{1.5ex plus .2ex}{\it }}
\makeatletter\renewcommand{\subsubsection}{\@startsection{subsubsection}{3}{-2.45ex}{-3.25ex
plus -1ex minus -.2ex}{1.5ex plus .2ex}{\it }}
\renewcommand{\thesection}{\arabic{section}}

\renewcommand{\theequation}{\thesection.\arabic{equation}}
\makeatletter \@addtoreset{equation}{section}

\usepackage{graphicx}
\usepackage{epsfig} 
\usepackage{verbatim}
\usepackage{graphicx}
\usepackage{epsfig} 
\usepackage{hyperref}
\usepackage{slashed}
\renewenvironment{thebibliography}[1]
     {\baselineskip=16pt plus 2pt minus 1pt
      \section*{\large\refname
        \@mkboth{\MakeUppercase\refname}{\MakeUppercase\refname}}%
     \list{\@biblabel{\@arabic\c@enumiv}}%
           {\settowidth\labelwidth{\@biblabel{#1}}%
            \leftmargin\labelwidth
            \advance\leftmargin\labelsep
            \@openbib@code
            \usecounter{enumiv}%
            \let\p@enumiv\@empty
            \renewcommand\theenumiv{\@arabic\c@enumiv}}%
      \sloppy
      \clubpenalty4000
      \@clubpenalty \clubpenalty
      \widowpenalty4000%
      \sfcode`\.\@m}

\let\fn\footnote
\renewcommand{\footnote}[1]{\linespread{1.1}\fn{#1}\linespread{1.29}}

\usepackage{amsfonts}
\usepackage{amssymb}
\usepackage{mathrsfs}
\usepackage{amsmath,amssymb}
\usepackage{bm}
\usepackage{caption}
\newcommand{\appendices}{\section*{Appendix}\setcounter{section}{0} \setcounter{equation}{0}
\renewcommand{\thesection}{\Alph{section}.}
\renewcommand{\theequation}{\thesection\arabic{equation}}}

\def\tyng(#1){\hbox{\tiny$\yng(#1)$}}

\newcommand{\del}{\partial}
\def\I{{\mathbb I}}

\newcommand{\be}{\begin{equation}}
\newcommand{\ee}{\end{equation}}
\newcommand{\bea}{\begin{array}}
\newcommand{\ea}{\end{array}}
\newcommand{\beqa}{\begin{eqnarray}}
\newcommand{\eeqa}{\end{eqnarray}}
\newcommand{\nn}{\nonumber}

\begin{document}
\begin{titlepage}
\begin{flushright}
\end{flushright}

\vskip 2 em

\begin{center}
\centerline{{\Large \bf Quantum Hall Effect on Odd Spheres}} 


\vskip 5em

\centerline{\large \bf \"{U}.H. Co\c{s}kun, S. K\"{u}rk\c{c}\"{u}o\v{g}lu and G.C.Toga}

\vskip 1em

\centerline{\sl $^\dagger$ Middle East Technical University, Department of Physics,}
\centerline{\sl Dumlupinar Boulevard, 06800, Ankara, Turkey}

\vskip 1em

{\sl cumit@metu.edu.tr \,,} {\sl kseckin@metu.edu.tr\,,} {\sl toga.can@metu.edu.tr}

\end{center}
                                                                                                                                                                                                                                                                                                                                                                                                                                                                                                                                                                                                                                                                                                                                                                                                                                                                                                                                    
\vskip 4 em

\begin{quote}
\begin{center}
{\bf Abstract}
\end{center}

\vskip 1em

We solve the Landau problem for charged particles on odd-dimensional spheres $S^{2k-1}$ in the background of constant $SO(2k-1)$ gauge fields carrying the irreducible representation $\left ( \frac{I}{2}, \frac{I}{2}, \cdots, \frac{I}{2} \right)$. We determine the spectrum of the Hamiltonian, the degeneracy of the Landau levels and give the eigenstates in terms of the Wigner ${\cal D}$-functions, and for odd values of $I$ the explicit local form of the wave functions in the lowest Landau level (LLL). Spectrum of the Dirac operator on $S^{2k-1}$ in the same gauge field background together with its degeneracies is also determined and in particular the number of zero modes is found. We show how the essential differential geometric structure of the Landau problem on the equatorial $S^{2k-2}$ is captured by constructing the relevant projective modules. For the Landau problem on $S^5$, we demonstrate an exact correspondence between the union of Hilbert spaces of LLL's with $I$ ranging from $0$ to $I_{max} = 2K$ or $I_{max} = 2K+1$ to the Hilbert spaces of the fuzzy ${\mathbb C}P^3$ or that of winding number $\pm1$ line bundles over ${\mathbb C}P^3$ at level $K$, respectively. 

\vskip 1em

\vskip 5pt

\end{quote}

\end{titlepage}

\setcounter{footnote}{0}

\newpage

\section{Introduction}

In a recent article \cite{Hasebe:2014nia}, relation between the A-class topological insulators (TIs) and the quantum Hall effect (QHE) on even dimensional spaces has been explored and it has been recognized that A-class TIs can be realized as QHE in even dimensions \cite{Hasebe:2014nia, Hasebe:2003gx}. A-class TIs are not time-reversal invariant, appear in even dimension and can be charaterized via an integer topological invariant, while AIII-class are also not time reversal invariant, carry an integer topological invariant but appear in odd dimensions. In addition, AIII-class TIs have chiral symmetry, whereas the A-class TIs do not \cite{Ryu:2010zza}. Focusing on these connections between the A-class TIs and AIII-class TIs, in a subsequent article Hasebe \cite{Hasebe:2014eha} considered the possibility of realizing the latter type in terms of a quantum Hall system in odd dimensions. Elaborating on the formulation of QHE on the three sphere $S^3$, given by Nair \& Daemi\cite{Nair:2003st}, Hasebe found that Nambu 3-algebraic geometry is crucial for realizing the chirial symmetry of the TI in this setting and modelled the chiral TI as a superposition of two three spheres  embedded in $S^4$ with the $SU(2)$ background monopole fluxes, i.e. in the four-dimensional QHE of Hu and Zhang \cite{Zhang:2001xs}.   

Motivated by these developments, in this paper our aim is to investigate QHE on all odd-dimensional spheres $S^{2k-1}$. As we have already noted, QHE problem on $S^3$ is solved by Nair \& Daemi \cite{Nair:2003st} and a complementary treatment is recently given in Hasebe's work \cite{Hasebe:2014eha}\footnote{Other recent developments in solving Landau problem and Dirac-Landau problem in flat higher dimensional spaces are reported in \cite{Li1,Li:2012xja,Li:2011it}.}. The clear path for the construction of QHE over compact higher dimensional manifolds appear to be closely linked to the coset space realization of such spaces. Indeed odd spheres can also be realized as coset manifolds as $S^{2k-1}\equiv\frac{SO(2k)}{SO(2k-1)}\equiv\frac{Spin(2k)}{Spin(2k-1)}$. In their approach Nair \& Daemi took advantage of the fact that $S^3$ can also be realized as $S^3\equiv\frac{SU(2)\times SU(2)}{SU(2)_D}$ owing to the isomorphisms $\frac{SU(2)\times SU(2)}{\mathbf{Z}_2}=SO(4)$ and $\frac{SU(2)}{\mathbf{{Z}}_2}=SO(3)$, and they subsequently constructed the Landau problem for a charged particle on $S^3$ under the influence of a constant $SU(2)_D$ gauge field background carrying an irreducible representation (IRR) of the latter.
This quick approach is not immediately applicable to higher dimensional odd spheres. Nevertheless, coset space generalization of $S^{2k-1}$ in terms of the $SO(2k+1)$ can be used to handle this problem. 

A brief summary of our results and their organization in the present article is in order. In section 2, we set up and solve the Landau problem for charged particles on odd-dimensional spheres $S^{2k-1}$ in the background of constant $SO(2k-1)$ gauge fields carrying the irreducible representation $\left ( \frac{I}{2}, \frac{I}{2}, \cdots, \frac{I}{2} \right)$. In particular, we determine the spectrum of the Hamiltonian, the degeneracy of the Landau levels and give the eigenstates in terms of the Wigner ${\cal D}$-functions, and for odd values of $I$ the explicit local form of the wave functions in the lowest Landau level 
In this section, we demonstrate in detail how the essential differential geometric structure of the Landau problem on the equatorial $S^{2k-2}$ is captured by constructing the relevant projective modules and the related $SO(2k-2)$ valued curvature two-forms. We illustrate our general results on the examples of $S^3$ and $S^5$ for concreteness and in the latter case identify an exact correspondence between the union of Hilbert spaces of LLL's with $I$ ranging from $0$ to $I_{max} = 2K$ or $I_{max} = 2K+1$ to the Hilbert spaces of the fuzzy ${\mathbb C}P^3$ or that of winding number $\pm1$ line bundles over ${\mathbb C}P^3$ at level $K$, respectively. In section $3$ we determine the spectrum of the Dirac operator on $S^{2k-1}$ in the same gauge field background together with its degeneracies and also compute the number of its zero modes. Some relevant formulas from the representation theory of groups is given in a short appendix for completeness.

\section{Landau Problem on Odd Spheres $S^{2k-1}$}

\subsection{Basic set-up and the solution}

In this section we aim to set up and solve the Schr\"{o}dinger equation for charged particles on odd spheres, $S^{2k-1}$, under the influence of a constant background field. We will give the spectrum of the appropriate Hamiltonian for the problem and determine the associated wave-functions. In order to pose the problem in sufficient detail we start with laying out some definitions and conventions that are going to be used throughout the paper.

A convenient way of specifying the coordinates on $S^{2k-1}$ is to embed it in $\mathbb{R}^{2k}$. Then, $X_a \in \mathbb{R}^{2n}$, $a=(1,2,\dots,2k)$, satisfying the condition $X_{a}X_{a}=R^2$ gives the coordinates of $S^{2k-1}$ with radius $R$. Splitting of $X_{a}$ into certain spinorial coordinates is going to of essential interest in what follows. To see how this comes about, let us first note the well-known fact that the odd-dimensional spheres can be represented as the coset spaces
\be
S^{2k-1}=SO(2k)/SO(2k-1) \,,
\ee
and the generators of $SO(2k) \approx Spin(2k)$ may be given by 
\be 
\Xi_{ab} = - \frac{i}{4} \lbrack \Gamma_a \,, \Gamma_b \rbrack \,, \quad a,b = 1,2,\dots,2k
\label{SO2kgen}
\ee
where $\Gamma_a$ are the generators of the Clifford algebra in $2k$-dimensions. These are $2^k \times 2^k$ matrices satisfying the anti-commutation relations $\lbrace \Gamma_a \,, \Gamma_b \rbrace = 2 \delta_{ab}$.
We will use the following representation of $\Gamma_a$s in the present article:
\begin{align}
\Gamma_\mu &= \left(%
\begin{array}{cc}
0 & -i\gamma_\mu \\
i \gamma_\mu & 0 \\
\end{array}%
\right), \;\;\; \mu =1, \dots, 2k-1
\nonumber\\
\Gamma_{2k} &= \left(%
\begin{array}{cc}
0 & 1_{2^{k-1} \times 2^{k-1}} \\
1_{2^{k-1}\times 2^{k-1}} & 0 \\
\end{array}%
\right) \,, \quad \Gamma_{2k+1} = \left(%
\begin{array}{cc}
-1_{2^{k-1} \times 2^{k-1}} & 0 \\
0 & 1_{2^{k-1} \times 2^{k-1}} \\
\end{array}%
\right) \,,
\end{align}
where $\gamma_\mu$'s are the generators of the Clifford algebra in $(2k-1)$-dimensions. 

$SO(2k-1)$ is irreducibly generated by 
\be
\Sigma_{\mu \nu} = - \frac{i}{4} \lbrack \gamma_\mu \,, \gamma_\nu \rbrack \,.
\ee
\label{fspinor}
In fact, $\Sigma_{\mu \nu}$ specify the irreducible fundamental spinor representation of $SO(2k-1)$, which is $2^{k-1}$-dimensional. For completeness, let us also indicate that $\Xi_{ab}$ in \eqref{SO2kgen} generates $SO(2k)$ reducibly; 
it has the block diagonal form
\be
\Xi_{ab} = \left ( 
\begin{array}{cc}
\Xi_{ab}^+ & 0 \\
0 & \Xi_{ab}^-
\end{array}
\right )
\ee
indicating that there are two irreducible fundamental representations, $\Xi_{ab}^\pm = (\Xi_{\mu \nu}^\pm \,, \Xi_{2k \mu}^\pm ) = (\Sigma_{\mu \nu}, \mp \frac{1}{2} \gamma_\mu)$, each of dimension $2^{k-1}$, generating $SO(2k)$.

Let us introduce the $2^k$-component spinor 
\be
\Psi = \frac{1}{\sqrt{2R(R+x_{2k})}}((R+x_{2k}) \mathbb{I}_{2^k} + X_{\mu}\Gamma^{\mu}) \phi \,, \quad \quad \Psi^\dagger \Psi = 1\,,
\label{spinor1}
\ee
where $\mathbb{I}_{2^{k}}$ stands for a $2^{k}\times2^{k}$ unit matrix and $\phi =  \frac{1}{\sqrt{2}} \left(\begin{smallmatrix} {\tilde \phi}  \\ {\tilde \phi} \end{smallmatrix}\right)$, with ${\tilde \phi}$ being a normalized $2^{k-1}$-component spinor. It is straightforward to check that $\Psi$ gives us the desired fractionalization or the ``square root" of $x_a$ via the Hopf-like projection map 
\be
\frac{X_{a}}{R} = \Psi^{\dagger}\Gamma_{a}\Psi \,.
\ee 
Using the spinor introduced in \eqref{spinor1}, we can construct the spin connection over $S^{2k-1}$, i.e. the $SO(2k-1)$ gauge field as 
\be
A = \Psi^{\dagger} d \Psi \,,
\ee
whose components are determined to be
\be
A_\mu = -\frac{1}{R(R+X_{2k})}\Sigma_{\mu \nu}X_{\nu} \,, \quad A_{2k} = 0 \,.
\label{gaugecomp}
\ee
Using the covariant derivatives $D_{a} = \del_{a} + iA_{a}$ and \eqref{gaugecomp}, components of the field strength 
\be
F_{ab} = -i [D_a, D_b] = \partial_a A_b - \partial_b A_a + i \lbrack A_a, A_b \rbrack \,,
\ee
are given as
\be
F_{\mu \nu} = \frac{1}{R^2}(X_\nu A_\mu - X_\mu A_\nu + \Sigma_{\mu \nu})\,, \quad F_{2k\mu} = -\frac{R+X_{2k}}{R^2}A_\mu \,.
\label{conn11}
\ee
We find that
\be
R^{4} \sum_{a<b} F_{ab}^{2} = \sum_{\mu < \nu} \Sigma_{\mu\nu}^2 \,.
\label{background}
\ee
R.h.s of \eqref{background} is the Casimir of $SO(2k-1)$ and thus proportional to identity in an irreducible representation. Thus a natural choice for a constant gauge field
background is the spinor representation given by the highest weight labels \cite{Iachello} 
\be
\left( \frac{I}{2} \right)\equiv\underbrace{\left(\frac{I}{2},...,\frac{I}{2}\right)}_{(k-1) \text{ terms}} \,, \quad I \in {\mathbb Z} \,,
\label{IRR1}
\ee
since $SO(2k-1)$ is of rank $k-1$.  We observe that $\left(\frac{I}{2},...,\frac{I}{2}\right)$ can be obtained from the $I$-fold symmetric tensor product of the fundamental spinor representation $\left(\frac{1}{2},...,\frac{1}{2}\right)$. It should readily be understood from the context, which IRR of $SO(2k-1)$ that $\Sigma_{\mu\nu}$ carries; thus in \eqref{fspinor} this is the $2^{k-1}$-dimensional fundamental spinor representation $\left(\frac{1}{2},...,\frac{1}{2}\right)$, while in what follows we are going to take it to be in the IRR $\left(\frac{I}{2},...,\frac{I}{2}\right)$ due to the reasons just argued.     

We can write down the Hamiltonian for a charge particle on $S^{2k-1}$ under the influence of the constant $SO(2k-1)$ gauge field background introduced in the preceding paragraph as  
\be
H=\frac{\hbar}{2MR^{2}}\sum_{a<b}\Lambda_{ab}^{2} \,,
\label{Ham}
\ee
where $\Lambda_{ab}$ are the operators given as 
\be
\Lambda_{ab}= - i (X_a D_b - X_b D_a) \,,
\ee
which are parallel to the tangent bundle over $S^{2k-1}$. Commutators of $\Lambda_{ab}$ give 
\begin{multline}
\lbrack \Lambda_{ab}, \Lambda_{cd} \rbrack = i(\delta_{ac} \Lambda_{bd} + \delta_{bd}\Lambda_{ac} -
\delta_{bc} \Lambda_{ad} - \delta_{ad} \Lambda_{bc}) \\ 
- i (X_{a}X_{c} F_{bd} + X_{b}X_{d}F_{ac} - X_{b}X_{c}F_{ad} - X_{a}X_{d} F_{bc}) \,,
\end{multline}
which are not the $SO(2k)$ commutation relations. The reason for $\Lambda_{ab}$ failing to satisfy the $SO(2k)$ commutation relations is that, they just account for the angular momentum of a charged particle on $S^{2k-1}$, which is not the total angular momentum in the present problem since the background gauge field also carries angular momentum.
Thus, the total angular momentum operators, generating the $SO(2k)$ rotations can be constructed by supplementing $\Lambda_{ab}$ with the spin angular momentum of the background gauge field by writing
\be
L_{ab} = \Lambda_{ab} + R^2 F_{ab} \,,
\label{totalang}
\ee
In component form we find
\be
L_{\mu \nu} = L_{\mu \nu}^{(0)} + \Sigma_{\mu \nu} \,, \quad L_{2k \mu} =  L_{2k \mu}^{(0)} - R A_\mu \,,
\label{Lop}
\ee
where $L_{ab}^{(0)} = - i (X_a \partial_b - X_b \partial_a)$ are the generators of $SO(2k)$ over $S^{2k-1}$.  In the absence of a magnetic background $L_{ab}^{(0)}$ would be the generators of angular momentum for a particle on $S^{2k-1}$ and it would be the total angular momentum in that case. In the present case, a straightforward calculation yields
\be
\lbrack L_{ab}, L_{cd} \rbrack = i(\delta_{ac} L_{bd} + \delta_{bd}L_{ac} - \delta_{bc} L_{ad} - \delta_{ad} L_{bc}) \,,
\ee
as expected.

Using \eqref{totalang} and the fact that $\Lambda_{ab}$ and $F_{ab}$ are orthogonal, i.e. $\Lambda_{ab} F_{ab} = F_{ab} \Lambda_{ab} = 0$, we can write \eqref{Ham} as 
\be
H = \frac{\hbar}{2MR^{2}}\left(\sum_{a<b}L_{ab}^{2}-\sum_{\mu<\nu}\Sigma_{\mu \nu}^{2}\right) \,.
\label{Ham1}
\ee
In order to obtain the spectrum of this Hamiltonian, we have to determine the general form of the IRR of $SO(2k)$ that $L_{ab}$ could carry given that $\Sigma_{\mu\nu}$ carries the $\left ( \frac{I}{2} \right )$ of $SO(2k-1)$. This problem can be addressed by looking at the branching of $SO(2k)$ IRRs in terms of those of $SO(2k-1)$. Consider table 1
\begin{table}[htp]
\begin{center}
\begin{tabular}{ll} 
$SO(2k)$ & $\lambda_1 \,, \lambda_2 \,, \cdots \,, \lambda_{k-1} \,, \lambda_k$  \\ 
$SO(2k-1)$ & \, \, $\mu_1 \,, \, \, \mu_2 \,, \cdots \,, \, \, \mu_{k-1}$ \\
&$\lambda_1 \geq \mu_1 \geq \lambda_2 \cdots \geq \mu_{k-1} \geq |\lambda_k|$
\end{tabular} 
\caption{Branching of $SO(2k)$ under $SO(2k-1)$ }
\end{center}
\label{table1}
\end{table}
where first row in indicates a generic IRR of $SO(2k)$ labeled by integers or half odd integers $(\lambda_1, \lambda_2 \,,\cdots \,, \lambda_k)$ corresponding respectively to tensor and spinor representations with $\lambda_1 \geq \lambda_2 \cdots \geq |\lambda_k|$, where the last entry $\lambda_k$ could be positive, negative or zero and satisfies $|\lambda_k | \geq 0$ for the former and $|\lambda_k| \geq  \frac{1}{2}$ for the latter case. IRRs of $SO(2k)$ with opposite sign of $\lambda_k$ are conjugate representations. Second row stands for the $\left (\mu_1 \,, \mu_2 \,, \cdots \,, \mu_{k-1} \right )$ IRR of $SO(2k-1)$ where $\mu_i$ are integers or half odd integers satisfying $\mu_1 \geq \mu_2 \cdots \geq |\mu_{k-1} |$ and $\mu_{k-1} \geq 0$ or $\mu_{k-1} \geq \frac{1}{2}$, respectively. Third line gives the branching rule \cite{Iachello}. Accordingly, for $(\frac{I}{2})$ of $SO(2k-1)$ to appear in this branching, we must have $\lambda_1 \geq \frac{I}{2}$, $\lambda_2 = \lambda_3 = \cdots = \lambda_{k-1} = \frac{I}{2}$ and $|\lambda_k| \leq \frac{I}{2}$. Thus, we may write $\lambda_1 = n +  \frac{I}{2}$ for some integer $n$ and using the notation $\lambda_k = s$ ($s \leq \frac{I}{2}$) we see that $( n + \frac{I}{2}, \frac{I}{2}, \cdots, \frac{I}{2}, s)$ is the general form of the $SO(2k)$ IRR, whose branching under $SO(2k-1)$ includes the $\left (\frac{I}{2} \right )$ IRR of the latter. In fact the complete branching of the former can be written out as the direct sum of $SO(2k-1)$ IRRs as
\be
\left ( n + \frac{I}{2}, \frac{I}{2}, \cdots, \frac{I}{2}, s \right) = {\bigoplus}_{\mu_1= \frac{I}{2}}^{n+\frac{I}{2}} {\bigoplus}_{\mu_2=s}^{\frac{I}{2}} \left (\mu_1, \frac{I}{2}, \cdots, \frac{I}{2} ,\mu_2 \right ) \,.
\label{brrule}
\ee
Spectrum of the Hamiltonian can be written out using the eigenvalues of the quadratic Casimir operators $C^2_{SO(2k)}$ and $C^2_{SO(2k-1)}$ of $SO(2k)$ and $SO(2k-1)$ in the IRRs
$( n + \frac{I}{2}, \frac{I}{2}, \cdots, \frac{I}{2}, s)$, $(\frac{I}{2}, \frac{I}{2}, \cdots, \frac{I}{2})$, respectively. Eigenvalues for these Casimir operators in generic IRRs 
are given in the appendix. Explicitly, we have
\beqa
E &=& \frac{\hbar}{2MR^{2}}\left (C^2_{SO(2k)} \left( n + \frac{I}{2}, \frac{I}{2}, \cdots, \frac{I}{2}, s \right ) - C^2_{SO(2k-1)} \left (\frac{I}{2}, \frac{I}{2}, \cdots, \frac{I}{2} \right ) \right) \nn \\
&=& \frac{\hbar}{2MR^{2}} \left ( n^2 + s^2 + n(I+2k-2)+ \frac{I}{2}(k-1) \right ) \,.
\label{EnergySpectrum}
\eeqa
Thus, given a fixed background charge $I$, the lowest Landau level (LLL) is characterized by setting $n=0$ and $s=0$ if $I$ is an even integer or setting $n=0$ and $s = \pm \frac{1}{2}$ if $I$ is an odd integer. In these cases we get
\beqa 
E_{LLL} = 
\begin{cases}
\frac{\hbar}{2MR^{2}} \frac{I}{2}(k-1) & \mbox{for even $I$} \,, \\
\frac{\hbar}{2MR^{2}} \left ( \frac{I}{2}(k-1) + \frac{1}{4} \right) & \mbox{for odd $I$}
\end{cases}
\label{LLLEnergy}
\eeqa
It is possible to interpret $n$ and $s$ as the quantum numbers labeling the Landau levels. We further see that the degeneracy in each Landau level is given by the dimension of the
IRR $(n + \frac{I}{2}, \frac{I}{2}, \cdots, \frac{I}{2}, s)$ of $SO(2k)$, which can be written compactly as
\be
d(n,s)= \prod_{i<j}^k \left ( \frac{m_{i}-m_{j}}{g_{i}-g_{j}} \right )  \prod_{i<j}^k \left(\frac{m_{i}+m_{j}}{g_{i}+g_{j}} \right ) \,.
\label{deg1}
\ee
where $g_{i} = k - i$ and $m_1 = n + \frac{I}{2} + g_{1}$ \,, $m_i =\frac{I}{2} + g_{i}$ $(i=2,\cdots \,,k-1)$ and $m_k= s + g_k$. It is easy to estimate from \eqref{deg1} that for large $I$, $d(0,0) \approx I^{\frac{1}{2} (k-1)(k+2)} \approx d(0,\pm\frac{1}{2})$, and shows us how fast the LLL degeneracy grows for a given magnetic background on $S^{2k-1}$. We note also that for the LLL with $I$ odd the degeneracy is doubled since $s$ takes on the values $\pm \frac{1}{2}$.

In the thermodynamic limit $I, R \longrightarrow\infty$ with a finite ``magnetic length" scale  $ \ell_M = \frac{R}{\sqrt{I}} $, we immediately find
\be
E(n,s)\longrightarrow\frac{\hbar}{2M \ell_M^2} \left( n+\frac{1}{2}(k-1) \right) \,, \quad E_{LLL}=\frac{\hbar}{2M \ell_M^2} \frac{k-1}{2} \,,
\ee
and see that the spacing between LL levels remains finite and the LLL energy has the same form as in the standard IQHE in two dimensions up to an overall constant.

What about the wave functions corresponding to these Landau levels? Compactly they can be given in terms of the Wigner D-functions $D^{(n + \frac{I}{2}, \frac{I}{2}, \cdots, \frac{I}{2}, s)}(g)_{[L][R]}$ of $SO(2k)$ carrying the $(n + \frac{I}{2}, \frac{I}{2}, \cdots, \frac{I}{2}, s)$ IRR of the latter. Here $[L][R]$ are two sets of collective labels that give the states in this IRR of $SO(2k)$ with respect to the IRRs of $SO(2k-1)$ that appear in the branching \eqref{brrule}. Since $[R]$ labels all the states in the IRR $\left ( \frac{I}{2} \,, \cdots \,, \frac{I}{2}\right)$, $[L]$ is further subject to certain selection rules that restrict both the IRRs in \eqref{brrule} and states in each of the latter, which we do not attempt to determine here. Nevertheless, the $2^{k-1}$-component spinors
\be
\Psi^{\pm} = \frac{1}{2} \frac{1}{\sqrt{R(R+x_{2k})}}((R+x_{2k}) \mathbb{I}_{2^{k-1}} \mp i x_{\mu}\gamma^{\mu}) {\tilde \phi} \,, \quad \quad 
\Psi = \left (
\begin{array}{c}
\Psi^+ \\
\Psi^-
\end{array}
\right )
\ee
obtained from \eqref{spinor1} are indeed the LLL wave functions for $I =1$, with $\pm$ signs corresponding to $s= \frac{1}{2}$ and $s= -\frac{1}{2}$, respectively and $2^{k-1}$ fold degeneracy in each sector. Using the compact notation, $\Psi^\pm_\alpha := K_{\alpha \beta}^\pm {\tilde \phi}_\beta$, we see that
\be
L_{\mu \nu} \Psi^\pm_\alpha = K_{\alpha \beta}^\pm (\Sigma_{\mu \nu})_{\beta \gamma} {\tilde \phi}_\gamma \,, \quad L_{2k \mu} \Psi^\pm_\alpha =  K_{\alpha \beta}^\pm (\mp \frac{1}{2} \gamma_\mu)_{\beta \gamma} {\tilde \phi}_\gamma\,,
\ee
from which, after several steps of calculation, we find
\beqa
\sum_{a < b} L_{ab}^2 \Psi^\pm &=& \sum_{\mu < \nu} \left (\Sigma_{\mu \nu}^2 + \frac{1}{4}\gamma^2_\mu \right ) \Psi^\pm \,, \nn  \\
&=& \sum_{\mu < \nu} \Sigma_{\mu \nu}^2 + \frac{1}{2} \left ( k - \frac{1}{2} \right ) \,,
\eeqa
indicating the claimed result upon using \eqref{Ham1}. Thus the LLL wave functions for the case of odd $I$ are then obtained by the $I$-fold symmetric product of $\Psi^\pm_\alpha$
\be
\Psi^I = \sum_{\alpha_1\,, \cdots \alpha_I}  f_{\alpha_1\, \cdots \alpha_I} \, \Psi_{\alpha_1} \cdots \Psi_{\alpha_I} \,,
\ee
where each $\alpha$ takes on values from 1 to $2^{k-1}$ and the coefficients $f_{\alpha_1\, \cdots, \alpha_I}$ are totally symmetric in its indices and satisfy $\Gamma^a_{\alpha_1 \alpha_2} f_{\alpha_1\,  \alpha_2 \cdots \alpha_I} = 0$, $f_{\alpha \, \alpha \alpha_3 \cdots \alpha_I} = 0$ to exclude the non-symmetric representations that appear in the $I$-fold tensor product of $\left (\frac{1}{2} \right)$ IRR of $SO(2k-1)$.

For $N$ particles the LLL wave function can be obtained via the Slater determinant of $\Psi_I$ and reads
\be
\Psi^I_N = \sum_{\alpha_1\,, \cdots \alpha_I}  \varepsilon_{\alpha_1\, \cdots \alpha_I} \, \Psi^I_{\alpha_1}(x_1) \cdots \Psi^I_{\alpha_I}(x_N) \,,
\ee
where $\varepsilon_{\alpha_1\, \cdots, \alpha_I}$ is the usual permutation symbol, which is totally antisymmetric in its indices.

\subsection{The Equatorial $S^{2k-2}$}

It appears possible to probe further the physics at the equatorial spheres $S^{2k-2}$. To see how the physics matches with the known results of Landau problem on even spheres $S^{2k-2}$ we proceed as follows. We first note that
\be
(K^\pm)^2 = \frac{1}{R} (x_{2k}(\mathbb{I}_{2^{k-1}} \mp i x_{\mu}\gamma^{\mu}) \,, \quad (K^\pm_0)^2 := (K^\pm)^2 \Big |_{{x_{2k}} = 0} = \mp i \frac{1}{R} x_{\mu}\gamma^{\mu} \,.
\ee
We may now define the idempotent on the equatorial $S^{2k-2}$ as 
\be 
Q = i (K^\pm_0)^2 \,, \quad Q^\dagger = Q \,, \quad  Q^2 = \mathbb{I}_{2^{k-1}} \,,
\ee
which allows us to write down the rank $1$ projection operators 
\be
{\cal P}_{\pm} = \frac{\mathbb{I}_{2^{k-1}} \pm Q}{2} \,.
\ee
Denoting the algebra of functions on $S^{2k-2}$ as ${\cal A}$, we may write the free ${\cal A}$-module
as ${\cal A}^{2^{k-1}} = {\cal A} \otimes {\mathbb C}^{2^{k-1}}$ and form the projective modules ${\cal P}_{\pm} {\cal A}^{2^{k-1}}$. In other words, we may decompose the free ${\cal A}^{2^{k-1}}$-module as 
\be
{\cal A}^{2^{k-1}} = {\cal P}_+ {\cal A}^{2^{k-1}} \oplus {\cal P}_- {\cal A}^{2^{k-1}} \,,
\ee
where each summand is of dimension $2^{k-2}$.

Projections of rank $I$ are obtained by writing 
\be
{\cal P}_{\pm}^I = \prod_{i=1}^I \frac{\mathbb{I} \pm {\cal Q}_i}{2} \,, \quad {\cal Q}_i = \mathbb{I}_{2^{k-1}} \otimes \mathbb{I}_{2^{k-1}} \otimes \cdots \otimes Q \otimes \cdots \otimes \mathbb{I}_{2^{k-1}} \,,
\ee
where ${\cal Q}_i$ is an $I$-fold tensor product whose $i^{th}$ entry is $Q$. ${\cal P}_{\pm}^I$ and ${\cal Q}_i$ act on the free module ${\cal A}^{2^{k-1}}_I = {\cal A} \otimes {\mathbb C}^{2^{k-1}}_{I}$, where ${\mathbb C}^{2^{k-1}}_{I}$ is the $I$-fold symmetric tensor product of ${\mathbb C}^{2^{k-1}}$, whose dimension is that of the $\left (\frac{I}{2}, \frac{I}{2}, \cdots, \frac{I}{2} \right)$ IRR of $SO(2k-1)$

$SO(2k-1)$ and $SO(2k-2)$ are groups of rank $k-1$ and the branching of the $\left (\frac{I}{2}, \frac{I}{2}, \cdots, \frac{I}{2} \right)$ IRR of the former under the IRRs of the latter reads
\be
\left (\frac{I}{2}, \frac{I}{2}, \cdots, \frac{I}{2} \right) = \bigoplus_{\mu = - \frac{I}{2}}^{\frac{I}{2}} \left ( \frac{I}{2}, \frac{I}{2}, \cdots, \frac{I}{2}, \mu \right ) \,.
\label{sodec1}
\ee  
${\cal P}_{\pm}^I$ are indeed the projections to the $\left (\frac{I}{2}, \frac{I}{2}, \cdots, \pm \left | \frac{I}{2} \right | \right)$ IRRs of $SO(2k-2)$ appearing in the r.h.s. of the decomposition given in \eqref{sodec1}. These are the projective modules ${\cal P}_{\pm}^I {\cal A}^{2^{k-1}}_I$ whose dimensions are equal and given by the dimension of $\left (\frac{I}{2}, \frac{I}{2}, \cdots, \pm \left | \frac{I}{2} \right | \right)$.

We are now in a position to observe that the connection two-forms associated with ${\cal P}_{\pm}^I$ are \cite{FuzzyBook}
\be
{\cal F}_\pm = {\cal P}_{\pm}^I \, d \, ( {\cal P}_{\pm}^I) \, d \, ({\cal P}_{\pm}^I ) \,.
\ee
Thus it follows from the remark ensuing \eqref{sodec1} that ${\cal F}_\pm $ are nothing but the $SO(2k-2)$ constant background gauge fields on $S^{2k-2}$ which are characterized by the IRRs $\left (\frac{I}{2}, \frac{I}{2}, \cdots,  \pm \left | \frac{I}{2} \right | \right)$ of $SO(2k-2)$. Finally, we note that the $(k-1)^{th}$ Chern number is given by 
\be
c^\pm_{k-1} = \frac{1}{k! (2\pi)^k} \int _{S^{2k-2}} {\cal P}_{\pm}^I \left (d \, ( {\cal P}_{\pm}^I  ) \right)^{2k-2} \,.
\ee
where $c_{k-1}\equiv c^+_{k-1}>0$ and $c^-_{k-1}=-c_{k-1}$. These numbers match with the degeneracy of the the LLL on $S^{2k-2}$ via the relation $c_{k-1}(I) =d_{LLL}^{S^{2k-2}}(k-1, I-1)$. $c_{k-1}(I)$ also matches exactly with the number of zero modes, i.e. the index of the gauged Dirac operator on $S^{2k-2}$, as an independent solution of the Landau problem and Dirac-Landau problem on $S^{2k-2}$ given in \cite{Hasebe:2014eha} confirms. Our brief analysis in this subsection shed further light on the relation between QHE problem over even and odd spheres. 

\subsection{QHE on $S^3$} \label{qheons3}

This is the case considered first by Nair and Daemi \cite{Nair:2003st} and recently by Hasebe \cite{Hasebe:2014eha}. $S^3\equiv SO(4)/SO(3)$, which follows by setting $k=2$ in \eqref{EnergySpectrum}. Energy spectrum takes the form
\be \label{S3}
E = \frac{\hbar}{2MR^{2}}(n^2 + 2n + In + \frac{I}{2} + s^2)
\ee
and the degeneracy of (\ref{S3}) is given by the dimension of the $\left (n+ \frac{I}{2} \,, s \right)$ IRR of $SO(4)$
\be
d(n,s)= (n+\frac{I}{2}+s+1)(n+\frac{I}{2}-s+1) = (n+\frac{I}{2}+1)^2 -s^2
\ee 
For the LLL we have
\be
E_{LLL} = \frac{\hbar}{2MR^{2}} \frac{I}{2} \,, \quad \mbox {I even} \,, \quad \quad E_{LLL} = \frac{\hbar}{2MR^{2}} \left(\frac{I}{2} + \frac{1}{4} \right) \,, \quad  \mbox{I odd}
\ee
with the degeneracies 
\be
d(n=0,s=0)=(\frac{I}{2}+1)^2 \,, \quad d(n=0,s=\pm\frac{1}{2}) = d(0,+1/2) + d(0,-1/2) = \frac{1}{2}(I+1)(I+3) \,,
\ee
which are all in agreement with the results of \cite{Nair:2003st} and \cite{Hasebe:2014eha}.

\subsection{QHE on $S^5$}

Our next example is $S^5\equiv SO(6)/SO(5)$, which follows from setting $k=3$ in \eqref{EnergySpectrum}. In this case the energy spectrum takes the form
\be \label{S5}
E = \frac{\hbar}{2MR^{2}}(n^2 + 4n + In + I + s^2)
\ee
with the degeneracy of (\ref{S5}) given by the dimension of the $\left (n+ \frac{I}{2} \,, \frac{I}{2} \,, s \right)$ IRR of $SO(6)$ as
\be
d(n,s) =\frac{1}{12}(n+1)^2 (n+I+3) \left ((n+\frac{I}{2}+2)^2-s^2 \right) \left( (\frac{I}{2}+1)^2- s^2 \right) \,.
\ee
Inspecting the LLL, we can write down the energy spectrum and degeneracies as 
\be
E_{LLL}=\frac{\hbar}{2MR^{2}}I \,, \quad \mbox {I even} \,, \quad \quad E_{LLL} = \frac{\hbar}{2MR^{2}} \left(I + \frac{1}{4} \right) \,, \quad  \mbox{I odd}
\ee
\be
d(n=0,s=0)=\frac{1}{3 \cdot 2^6}(I+2)^{2}(I+3)(I+4)^{2} \,, \quad \mbox {I even} \,,
\ee 
and 
\be
d(n=0,s=\pm\frac{1}{2})=d(0,+1/2)+d(0,-1/2)=\frac{1}{3 \cdot 2^5}(I+1)(I+3)^3(I+5) \,, \quad \mbox {I odd} \,.
\ee

There is an exact correspondence between the union of Hilbert spaces of LLLs with $I$ ranging from $0$ to $I_{max} = 2K$ or $I_{max} = 2K+1$ correspond respectively to the Hilbert spaces of the fuzzy ${\mathbb C}P^3$ or that of winding number $\pm1$ line bundle over ${\mathbb C}P^3$ at level $K$ \cite{FuzzyBook, Dolan:2003kq}. This interesting relationship essentially follows due to the fact that the isometry group $SU(4)$ for ${\mathbb C}P^3$ is isomorphic to that of $S^5$ which is $Spin(6) \approx SO(6)$. We can demonstrate this relation very easily.

Let us recall that the fuzzy ${\mathbb C}P^3$ at level $K$ is given in term of the matrix algebra $Mat(d_K)$, where $d_K = \frac{1}{6} (K+3)(K+2)(K+1)$. It covers all the IRRs of   
$SU(4)$ which emerge from the tensor product 
\be
\left (\frac{K}{2}, \frac{K}{2}, \frac{K}{2} \right) \otimes \left (\frac{K}{2}, \frac{K}{2}, - \frac{K}{2} \right) = \bigoplus_{k=0}^K (k,k,0)
\label{cp3}
\ee
Expansion of an element of $Mat(d_K)$ in terms of $SU(4)$ harmonics carries the IRRs of $SU(4)$ appearing in the direct sum decomposition given in the r.h.s of \eqref{cp3}. We observe, that each summand in the latter is equal to the $SU(4) \approx SO(6)$ IRR carried by the LLL for $I=2k$. This readily implies that, for even $I$, $I =2k$, the union of all the LLL Hilbert spaces with $0 \leq k \leq 2K$ spans the matrix algebra $Mat(d_K)$ of ${\mathbb C}P^3_F$. 

Sections of complex line bundles with winding number $1$ over ${\mathbb C}P^3_F$ are described via the tensor product decomposition
\be
\left (\frac{K+1}{2}, \frac{K+1}{2}, \frac{K+1}{2} \right) \otimes \left (\frac{K}{2}, \frac{K}{2}, - \frac{K}{2} \right) = \bigoplus_{k=0}^K (k+\frac{1}{2},k+\frac{1}{2},\frac{1}{2})
\label{cp3sections}
\ee
Elements in this nontrivial line bundle are $d_{K+1} \times d_K$ rectangular matrices forming a right module ${\cal A}^{(1)}({\mathbb C}P^3_F)$ under the action of $Mat(d_K)$. We observe that each summand in the r.h.s of \eqref{cp3sections} corresponds to an $SO(6)$ IRR carried by the LLL for $I=2k+1$ ans $s = \frac{1}{2}$. Thus, the union of all the LLL Hilbert spaces with $0 \leq k \leq 2K+1$ spans ${\cal A}^{(1)}({\mathbb C}P^3_F)$ over ${\mathbb C}P^3_F$. In particular, it is straightforward to check that total number of states in this union of LLL is precisely $d_{K+1} d_K$:
\be
\sum_{k=0}^K \frac{1}{12} (k+4) (k+3) (k+2)^2 (k+1) = d_{K+1} d_K
\ee
A similar correspondence for the unions of LLLs with $s = - \frac{1}{2}$ and ${\cal A}^{-1}({\mathbb C}P^3_F)$ corresponding to winding number $-1$ sectors is established starting with the tensor product $\left (\frac{K}{2}, \frac{K}{2}, \frac{K}{2} \right) \otimes \left (\frac{K+1}{2}, \frac{K+1}{2}, - \frac{K+1}{2} \right)$.

\section{Dirac-Landau Problem on $S^{2k-1}$}

In this section, our aim is to determine the spectrum of the Dirac operator for charged particles on $S^{2k-1}$ under the influence of a constant $SO(2k-1)$ gauge field background.

Let us briefly recall the situation in the absence of a background gauge field. In this case Dirac operator for odd dimensional spheres $S^{2k-1}$ is well-known. It can be expressed in the form \cite{Balachandran:2002bj} 
\be
{\cal D}^{\pm} = \frac{1}{2}(\I \mp \Gamma_{2k+1}) \sum_{a<b} (-\Xi_{ab}L^{(0)}_{ab}+k-\frac{1}{2}) \,,
\ee
where $L_{ab}^{(0)}$ is given after $\eqref{Lop}$ and carriers the $(n,0,\cdots,0)$ IRR of $SO(2k)$ and $\Xi_{ab}$ given in \eqref{SO2kgen} carries the reducible representation $\left(\frac{1}{2},\frac{1}{2}, \cdots,\frac{1}{2}\right) \oplus \left(\frac{1}{2}, \frac{1}{2}, \cdots, -\frac{1}{2}\right)$ of $SO(2k)$. The projectors $ {\mathscr{P}}^{\mp} = \frac{1}{2}(\I \mp \Gamma_{2k+1})$ allows us to pick either the two inequivalent representations. To obtain the spectrum of ${\cal D}^{\pm}$, we simply need to observe that 
\be
(n,0,\cdots,0) \otimes \left(\frac{1}{2},\frac{1}{2}, \cdots, \pm \frac{1}{2}\right) = \left(n + \frac{1}{2},\frac{1}{2}, \cdots, \pm \frac{1}{2}\right) \oplus \left(n - \frac{1}{2},\frac{1}{2}, \cdots,\mp \frac{1}{2}\right) \,,
\label{spinorproduct1}
\ee
Since the $\left(\frac{1}{2},\frac{1}{2}, \cdots, \pm \frac{1}{2} \right)$ IRRs of $SO(2k)$ are conjugates, both representations yield the same spectrum for the Dirac operator ${\cal D}^{\pm}$ as expected, which is found to be \cite{Balachandran:2002bj}
\be
E_\uparrow = n + k -\frac{1}{2} \,, \quad E_\downarrow = - (n + k -\frac{3}{2}) \,,   
\label{diracspec1}
\ee
for the spin up and spin down states, respectively. Using the notation $j_{\uparrow \downarrow} = n \pm \frac{1}{2}$, we can express the spectrum of ${\cal D}^{\pm}$ more compactly as $E_{\uparrow \downarrow}  = \pm (j_{\uparrow \downarrow} + k - 1)$.

Let us now consider the gauged Dirac operator, which can be written by replacing ${\cal L}^{(0)}_{ab}$ with ${\Lambda}_{ab} = L_{ab} - R^2 F_{ab}$ as  
\be
{\cal D}_G^{\pm} = \frac{1}{2}(\I \mp \Gamma_{2k+1})  \sum_{a<b} \left (-\Xi_{ab}(L_{ab} - R^2 F_{ab})+k-\frac{1}{2} \right ) \ .
\ee
It is not possible to obtain the spectrum ${\cal D}_G$ in the same manner as that of the zero gauge field background case. There is, however, a well-known formula on symmetric spaces that relate the square of the gauged Dirac operator to the gauged Laplacian, the Ricci scalar of the manifold under consideration and a Zeeman energy term related to the curvature of the background gauge field \cite{Dolan:2003bj}. Furthermore, on a symmetric coset space, say $ K \equiv G\big/ H$, a particular gauge field background which is compatible with the isometries of $K$ generated by $G$ (in the sense that the Lie derivative of the gauge field strength along a Killing vector of $K$ is a gauge transformation of the field strength)   
is given by taking the gauge group as the holonomy group $H$ and identifying the gauge connection with the spin connection. Then the square of the Dirac operator can be expressed as \cite{Dolan:2003bj}
\be
(i {\cal D}_G^{\pm})^2 = C^2(G) - C^2(H) + \frac{\cal{R}}{8} \,,
\ee
where ${\cal R}$ is the Ricci scalar of the manifold $K$ and $C^2(G)$ and $C^2(H)$ are quadratic Casimirs of $G$, $H$, respectively, where $C^2(H)$ is evaluated in the IRR of $H$ characterizing the background gauge field, while $C^2(G)$ is evaluated in certain IRRs of $G$ containing the fixed combinations of the background isospin of the gauge field and the intrinsic spin of the fermion.  
These considerations fit perfectly with our problem for odd spheres $S^{2k-1}$ under fixed $SO(2k-1)$ gauge field backgrounds, since in the present problem we have taken the gauge group as the holonomy group $SO(2k-1)$ of the odd-spheres and the gauge connection is identified with the spin connection and taken explicitly in the IRR of $SO(2k-1)$ which is the $I$-fold symmetric tensor product of the fundamental spinor representation $\left (\frac{1}{2}\,,\cdots \frac{1}{2} \right)$. Therefore we can write     
\be
\label{diracsquare}
(i {\cal D}_G^{\pm})^2 
= C^2_{SO(2k)} \left( n + J , J, \cdots, J, \pm {\tilde s} \right ) - C^2_{SO(2k-1)} \left (\frac{I}{2}, \frac{I}{2}, \cdots, \frac{I}{2} \right ) + \frac{1}{4}(2k^2-3k+1)
\ee
where $2(2k^2-3k+1)$ is nothing but the Ricci scalar of the sphere $S^{2k-1}$ and $J$ takes on the values $J = \frac{I}{2} + \frac{1}{2}$ ($I\geq 0$) and $J = \frac{I}{2} - \frac{1}{2}$ ($I \geq 1$) corresponding to the spin up and spin down states, respectively and $|{\tilde s}| \leq J$. We find 
\beqa
\label{diracspecspinup}
{\cal E}_\uparrow &=& n(n+2k-1) + I(n+k-1) + k(k-1) + {\tilde s}^2 \,, \quad I \geq 0 \,, \\
{\cal E}_\downarrow &=& n(n+I+2k-3) + {\tilde s}^2 \,, \quad I \geq 1
\label{diracspecspindown}
\eeqa
It is readily seen that the spectrum of left and right chiral cases coincide with ${\tilde s} \rightarrow -{\tilde s}$. 

Degenarcy of ${\cal E}_\uparrow$ and ${\cal E}_\downarrow$ are given by the dimensions of the IRRs $\left( n + J , J, \cdots, J, s \pm \frac{1}{2} \right )$ with  $J = \frac{I}{2} + \frac{1}{2}$ and  $J = \frac{I}{2} - \frac{1}{2}$, respectively. They can be computed from \eqref{deg1} with $g_{i} = k - i$ and $m_1 = n + J + g_{1}$ \,, $m_i = J + g_{i}$ $(i=2,\cdots \,,k-1)$ and $m_k= {\tilde s} + g_k$. 

The Hamiltonian for the Dirac-Landau problem may be taken as $H = \frac{1}{2MR^2}(i {\cal D}_G^{\pm})^2$. For even $I$, we see that then the LLL is given by taking $n=0$ and ${\tilde s}= \pm \frac{1}{2}$ in \eqref{diracspecspindown} yielding ${\cal E}^{LLL}_\downarrow = \frac{1}{4}$ with the same degeneracy for both the operators given as $d(n = 0, {\tilde s}= \frac{1}{2}) + d(n=0 \,, {\tilde s} = -\frac{1}{2})$, which can be computed from \eqref{deg1} using the facts given in the previous paragraph. For odd $I$, we see that LLL is given by taking $n=0$ and ${\tilde s} =0 $ in \eqref{diracspecspindown} yielding ${\cal E}^{LLL}_\downarrow = 0$. These are the zero modes of the Dirac operators ${\cal D}_G^{\pm}$ with the degeneracy $d(n = 0, {\tilde s} = 0)$.
 
For $S^3$, we find that the LLL  degeneracy for even $I$ is given as $\frac{I(I+2)}{4}$ and for odd $I$ it is $\frac{(I+1)^2}{4}$, which is the number of zero modes of Dirac operators ${\cal D}_G^{\pm}$. These match with results of \cite{Nair:2003st}. Another example is $S^5$, with the LLL  degeneracy for even $I$ given as $\frac{1}{3 \cdot 2^6} I (I+2)^3 (I+4)$, and for odd $I$ it is $\frac{1}{3 \cdot 2^6} (I+1)^2 (I+2) (I+3)^2$. 

We may recall that on even dimensional manifolds, Atiyah-Singer index theorem relates the number of zero modes, i.e. index of the Dirac operator to Chern classes, which are integers of topological significance \cite{Eguchi:1980jx}. On odd dimensional manifolds, however, there is known such general index theorem. One possible candidate for a topological number on these manifolds could be conceived as the Chern-Simons forms. Nevertheless, for odd spheres it is not too hard to see that these vanish identically when evaluated for the $SO(2k-1)$ connection given in \eqref{conn11}. Thus, it remains an open question to find out if and how the zero modes of ${\cal D}_G^{\pm}$ are related to a number of topological origin.  

Finally, let us also note that setting $I =0$ in \eqref{diracspecspinup}, we have ${\tilde s}= \pm \frac{1}{2}$ and we find ${\cal E}_\uparrow = (n+k - \frac{1}{2})^2$, which matches with the known result for ${\cal D}^{\pm}$ given in \eqref{diracspec1}. Explicitly, we have $E_\uparrow = \sqrt{{\cal E}_\uparrow}$, while $E_\downarrow = - \sqrt{{\cal E}_\uparrow}$ with $n \rightarrow n-1$. The latter is necessary to match the IRR $\left( n + \frac{1}{2} , \frac{1}{2}, \cdots, \frac{1}{2}, \pm \frac{1}{2} \right )$ with the second summand in \eqref{spinorproduct1}. 

\vskip 2em

{\bf \large Acknowledgements}

This work is supported by the Middle East Technical University under Project No. BAP--01-05-2016-002.

\appendices

\section{Some Representation Theory}

\subsection{Branching Rules}

Irreducible representations of $SO(\cal N)$ and $SO({\cal N}-1)$ can be given in terms of the highest weight labels $[\lambda]\equiv(\lambda_1 \,, \lambda_2 \,, \cdots \,, \lambda_{{\cal N} -1 } \,, \lambda_{\cal N})$ and $[\mu]\equiv(\mu_1 \,, \mu_2 \,, \cdots \,, \, \,\mu_{{\cal N}-1})$ respectively. Branching of the IRR $[\lambda]$ of $SO(\cal N)$ under $SO({\cal N}-1)$ IRRs follows from the rule \cite{fh}

\be
[\lambda]= \bigoplus_{\lambda_1\ge \mu_1 \ge \lambda_2\ge \mu_2\ge \cdots \ge \lambda_{k-1}\ge \mu_{k-1}\ge |\lambda_k|} [\mu] , \hspace{1cm} \text{for} \hspace{0.75cm} {\cal N} = 2k
\ee

\be
[\lambda]= \bigoplus_{\lambda_1\ge \mu_1 \ge \lambda_2\ge \mu_2\ge \cdots \ge \lambda_{k-1}\ge \mu_{k-1}\ge \lambda_k\ge |\mu_k|} [\mu] , \hspace{1cm} \text{for} \hspace{0.75cm} {\cal N} = 2k + 1
\ee

\subsection{Quadratic Casimir operators of $SO(2k)$ and $SO(2k-1)$ Lie algebras}

Eigenvalues for the quadratic Casimir operators of $SO(2k)$ and $SO(2k-1)$ in the IRRs $[\lambda]\equiv(\lambda_1, \lambda_2 \,\cdots \, \lambda_k)$, $[\mu]\equiv(\mu_1, \mu_2 \,\cdots \, \mu_{k-1})$, respectively are given as \cite{Iachello}:
\beqa
C_2^{SO(2k)}[\lambda] = \sum_{i=1}^{k}\lambda_{i}(\lambda_{i}+2k-2i)\, \\
C_2^{SO(2k-1)}[\mu]= \sum_{i=1}^{k-1}\mu_{i}(\mu_{i}+2k-1-2i) \,.
\eeqa
Eigenvalues of quadratic Casimir operators of some specific IRRs are given as 
\beqa
C_2^{SO(4)}\left(n+\frac{I}{2},s\right)= \frac{I^2}{4}+I n+I+n^2+2 n+s^2  \\
C_2^{SO(3)}\left(\frac{I}{2}\right)=\frac{I^2}{4} + \frac{I}{2}\\
C_2^{SO(6)}\left(n+\frac{I}{2},\frac{I}{2},s\right)=\frac{I^2}{2}+I n+3 I+n^2+4 n+s^2 \\
C_2^{SO(5)}\left(\frac{I}{2},\frac{I}{2}\right)=\frac{I^2}{2} + 2I
\eeqa

\subsection{Relationship between Dynkin and Highest weight labels}

Throughout this paper highest weight labels (HW) have been used to label the irreducible representations of Lie algebras. Another common way to label the IRRs is given by the Dykin indices. The relationship between Dykin indices and highest weight labels are as follows.

For a $SO(4)$ IRR the labels are
\be
(p,q)_{Dynkin} \equiv (\lambda_1,\lambda_2)_{HW} \,,
\ee
where the relation between these labels are given by
\be 
p= (\lambda_1+\lambda_2)  \,, \hspace{1cm} q = (\lambda_1-\lambda_2) \,.
\ee
For instance, $\left (n+\frac{I}{2},s \right)_{HW}$ which is the IRR used in section \ref{qheons3} to label the LL on $S^3$ corresponds to $\left( n+\frac{I}{2}+s, n+\frac{I}{2}-s \right)_{Dynkin}$, whle the LLL are given by either $\left(\frac{I}{2}, \frac{I}{2} \right)_{Dynkin}$ or $\left( \frac{I}{2} \pm \frac{1}{2}, \frac{I}{2} \mp \frac{1}{2} \right)_{Dynkin}$.

For a $SO(5)$ IRRs, the labels are 
\be
(p,q)_{Dynkin} \equiv (\lambda_1,\lambda_2)_{HW} \,,
\ee
and the relation between these labels are given by
\be 
p = \lambda_1-\lambda_2  \,,  \hspace{1cm} q=2\lambda_2 \,,
\ee 
For instance, $I$-fold symmetric tensor product of $(\frac{1}{2},\frac{1}{2})_{HW}$ is $(\frac{I}{2},\frac{I}{2})_{HW}$ and in terms of Dynkin index labels this corresponds to $(0,I)_{Dynkin}$.

Finally, for $SO(6)$ IRRs the labels are given as
\be
(p,q,r)_{Dynkin} \equiv (\lambda_1,\lambda_2,\lambda_3)_{HW} \,,
\ee
and the relation between these labels are given by
\be 
p= \lambda_2 + \lambda_3 \hspace{1cm} q= \lambda_1 - \lambda_2 \hspace{1cm} r= \lambda_2 - \lambda_3 \,. 
\ee

\end{document}